\documentclass[lettersize,journal]{IEEEtran}
\usepackage[utf8]{inputenc}
\usepackage{amsmath,amssymb,amsfonts}
\usepackage{algorithmic}
\usepackage{graphicx}
\usepackage{textcomp}
\usepackage{xcolor}
\usepackage{multicol}
\usepackage{makeidx} 
\usepackage{amsthm}
\usepackage{algorithm}
\usepackage{relsize}
\usepackage{subfigure}
\usepackage{url}
\usepackage{cite}
\usepackage[english]{babel}
\usepackage{longtable}
\usepackage{tabularx}
\usepackage{ltablex}
\usepackage{array}
\usepackage{wrapfig}
\usepackage{tikz}
\usetikzlibrary{positioning}
\usepackage{calc}
\usepackage{multirow}
\usepackage{float}

\usetikzlibrary{decorations.markings}
\tikzstyle{vertex}=[circle, draw, inner sep=0pt, minimum size=4pt]
\usetikzlibrary{arrows,calc} 
\usetikzlibrary{shapes.geometric, arrows}
\usetikzlibrary{arrows,decorations.markings}
\usepackage[
  separate-uncertainty = true,
  multi-part-units = repeat
]{siunitx}

\usepackage{hyperref}

\usepackage{blindtext}
\usepackage{ragged2e}

\usepackage{mathrsfs}
\usepackage{verbatim}

\def \c {{\bf c}}
\def \u {{\bf u}}
\def \v {{\bf v}}

\DeclareUnicodeCharacter{2212}{-}

\newtheorem{example}{Example}

\newtheorem{note}{Note}

\begin{document}

\makeatletter
\newenvironment{breakablealgorithm}
  {
   \begin{center}
     \refstepcounter{algorithm}
     \hrule height.8pt depth0pt \kern2pt
     \renewcommand{\caption}[2][\relax]{
       {\raggedright\textbf{\fname@algorithm~\thealgorithm} ##2\par}%
       \ifx\relax##1\relax 
         \addcontentsline{loa}{algorithm}{\protect\numberline{\thealgorithm}##2}%
       \else 
         \addcontentsline{loa}{algorithm}{\protect\numberline{\thealgorithm}##1}%
       \fi
       \kern2pt\hrule\kern2pt
     }
  }{
     \kern2pt\hrule\relax
   \end{center}
  }
\makeatother

\title{On the Classification of Codes over Non-Unital Ring of Order $4$}


 \author{\IEEEauthorblockN{Sourav~Deb,~\IEEEmembership{Student Member,~IEEE}, Isha~Kikani,~and~Manish~K.~Gupta,~\IEEEmembership{Senior Member,~IEEE} 
 \thanks{All the authors are with Dhirubhai Ambani Institute of Information and Communication Technology, Gandhinagar- 382007, India, email: \href{mailto:sourav_deb@ieee.org}{sourav\_deb@ieee.org}, \href{mailto:ishakikani999@gmail.com}{ishakikani999@gmail.com}, \href{mailto:mankg@guptalab.org}{mankg@guptalab.org}}}
\\ \IEEEauthorblockA{}
}




\maketitle

\begin{abstract}
In the last 60 years coding theory has been studied a lot over finite fields $\mathbb{F}_q$ or commutative rings $\mathcal{R}$ with unity.
Although in $1993$, a study on the classification of the rings (not necessarily commutative or ring with unity) of order $p^2$ had been presented, the construction of codes over non-commutative rings or non-commutative non-unital rings surfaced merely two years ago. In this letter, we extend the diverse research on exploring the codes over the non-commutative and non-unital ring $E= \langle 2a=2b=0, a^2=a, b^2=b, ab=a, ba=b \rangle$ by presenting the classification of optimal and nice codes of length $n\leq7$ over $E$, along-with respective weight enumerators and complete weight enumerators. 
\end{abstract}


\section{Introduction}
\IEEEPARstart{V}{arious} constructions of codes over finite commutative rings with unity are studied rigorously over a long time.
On the other hand, limited approaches can be found for codes over the non-commutative settings.
The preliminary work on classifying and determining the structure of the rings of order $p^2$ was given by Benjamin Fine \cite{ref1} in $1993$, which contained the impacting result that there are precisely $11$ rings of order $p^2$ up to isomorphism for prime $p$.
Apart from the first four rings of order $4$, $E$ is the first non-commutative and non-unital ring with a characteristic of $2$.
Most of the literature is focused on the ring $E$ to build the arena for codes over non-unital rings.
Despite the negligible quantity of literature over the non-unital rings, promising aspects can be anticipated from the accomplishments so far. 
While the construction of LCD and ACD codes strengthens the possibility of exploring cryptographic ideas, the development of DNA codes also bifurcates the research interest in non-unital rings.      
This justifies the necessity of theoretical and appropriate engineering study over these rings, which is the main motivation of this work.  

The ring $E$ is considered in \cite{ref2,ref3} with a rigorous description of the ring structure.
Due to the exceptional association of the size of a code with its dual code over $E$ contributes to the shifted focus upon Quasi-self dual codes (QSD Codes) instead of the Self-dual codes the researchers.       
The QSD codes and Type IV codes that are characterized as QSD Codes with even Hamming weight, of length $n<7$ over $E$ are constructed and classified using the build-up construction in \cite{ref2}, which is further extended for length $n<12$ in \cite{ref3}.
 

However, research articles exist in the literature, where the non-unital rings of orders $4$ and $6$ are explored.   
In \cite{ref4, ref5}, the authors considered the different rings $H$ and $I$ respectively.
Using two different build-up constructions, ``Quasi Self Dual (QSD) codes'' of size $2^n$ where $n$ denotes the length of the code are presented recursively.
Utilizing the semi-local property of $H$, the authors have constructed Type IV codes with a simple characterization.
Similar work can be found to classify the QSD codes of length up to $6$ over $I$.
This exploration is further extended in \cite{ref6}, where new variants of the build-up constructions over the ring $I$ are proposed to construct QSD codes, Type IV codes, and quasi-Type IV codes QTIV codes (defined as QSD codes built upon even torsion code) of length $n=7$ and $8$ with different weight distributions.

As a different approach, the construction of QSD codes of length up to $8$ over non-unital rings of order $6$ is found in \cite{ref7}.
Two different representations of these rings are presented, and the notion of the unimodular lattice is introduced over these rings. 

In continuation to the codes over $E$, Shi, Minjia, et al. in \cite{ref8} proposed QSD and Type IV codes with better minimum distance using a different approach that is based on adjacency matrices linked to two-class association schemes that are Strongly Regular Graphs (SRG) and Double Regular Tournaments (DRT).
In \cite{ref9}, researchers are motivated towards LCD and more general ACD codes over $E$ and characterize free LCD codes regarding binary generator matrices. 
It can be noted that the LCD and ACD codes that are constructed in this article are defined utilizing only the left multiplication property of the ring $E$.

As a different perspective on constructing DNA codes over four alphabets, researchers presented QSD DNA codes with fixed $GC$-content constraint built from QSD codes over the rings $E$ and $F$ in \cite{ref10}. 

Directed by the advancements in this regard, we present a comprehensive analysis of codes of length $n\leq7$ over the ring $E$.
We consider the largest minimum distance (optimality), generator matrix, and different weight enumerators as the underlying factor in classifying the codes in this letter.
Moreover, the aspect of nice codes is added in the proposed analysis for codes of length up to $6$. 

The organization in this letter is as follows.
Section \ref{pre} deals with the insight to the ring $E$ and the notations primarily used in the rest of the paper.
In Section \ref{classification}, we introduce the algorithms that are proved to be a handful in computing and present an inequivalent classification of optimal codes and nice codes for $n\leq6$.
Similarly, for $n=7$, we propose the classification of optimal codes in the same section.
Section \ref{conclusion} concludes this letter. 
The compiled data can be found in \url{https://guptalab.org/codesnur4/}.


\section{Preliminaries}
\label{pre}
  This work concentrates on the codes over the ring $E$ with characteristic $2$.
Eventually, $E$ is a ring of order four without the multiplicative identity, which adds the non-unital characteristics.
Moreover, $E$ comes up to be also non-commutative from its structural representation.
In \cite{ref1}, the ring representation is introduced to observe the ring uniquely as $E= \langle 2a=2b=0, a^2=a, b^2=b, ab=a, ba=b \rangle$, where $a$ and $b$ are the generators of the ring.

From the ring representation of $E$, one can derive the addition and multiplication table for $E$ with $c=a+b$ as,

\begin{equation*}
    \begin{tabular}{|c|c|c|c|c|c|}
        \hline
        + & 0 & a & b & c \\
        \hline
        0 & 0 & a & b & c \\
        \hline
        a & a & 0 & c & b \\
        \hline
        b & b & c & 0 & a \\
        \hline
        c & c & b & a & 0 \\
        \hline
    \end{tabular}  \  \  \  \ \ \ \ \ \ \ \ \begin{tabular}{|c|c|c|c|c|c|}
        \hline
        $\times$ & 0 & a & b & c \\
        \hline
        0 & 0 & 0 & 0 & 0 \\
        \hline
        a & 0 & a & a & 0 \\
        \hline
        b & 0 & b & b & 0 \\
        \hline
        c & 0 & c & c & 0 \\
        \hline
    \end{tabular}
\end{equation*}

Interesting modeling of the elements of $E$ can be found in \cite{ref3} where

\begin{equation*} 
a=\left(
\begin{array}{cc}
    0 & 0 \\
    0 & 1  \\
\end{array} \right), \ \text{and} \ b=\left(
\begin{array}{cc}
    0 & 1 \\
    0 & 1  \\
\end{array} \right).
\end{equation*}

$E$ can be seen as a local ring containing the maximal ideal $J=\{0,c\}$.
Subsequently, the correspondence  can be established as $E/J \cong \mathbb{F}_2$ by defining the reduction map $\tau: E \rightarrow E/J$ with respect to modulo $J$ where $\tau(0)=\tau(c)=0$ and $\tau(a)=\tau(b)=1$.

The subset $\mathscr{C} \subseteq E^n$ of length $n$ is a code over $E$.
In this work, we consider a liner code over $E$ as the left-sided $E$-submodule of $E^n$.
The Hamming distance for two strings $\u, \v \in E^n$ and is defined as $d_H= |\{i\ |\ u_i \neq v_i, 1 \leq i \leq n \}|$.
The minimum Hamming Distance $d_{min}$ for a code $\mathscr{C}$ of length $n$, defined by,  $d_{min}=\min\{d_H (\u,\v) : \forall \ \u,\v \in \mathscr{C}, \u \neq \v \}$.
So by denoting $|\mathscr{C}|=M$, we represent a code over the ring $E$ as $\mathscr{C}(n,M,d_{min})$.
Consequently, we define the Hamming Weight as $w_H(\u-\v)=d_H(\u,\v)$ and we assume $w_H(0)=0$, $w_H(a)=w_H(b)=w_H(c)=1$ in this work.
For any $\u= (u_1 u_2 \ldots u_n)$ for $u_i \in E, 1 \leq i \leq n$, the Hamming weight of a vector is stated by, $w_H(\u)=\sum\limits_{i=1}^n w_H(u_i)$.

Several weight enumerators are well defined in existing literature for a code over $\mathbb{Z}_4$.
In a similar manner, we define \textit{weight enumerator} or \textit{weight distribution} for an $E$-linear code $\mathscr{C}$ as $W_{\mathscr{C}}(z)= \sum\limits_{i=1}^n A_i.z^i$, where $A_i$ denotes the number of occurrence of codewords of weight $i$ in $\mathscr{C}$.
Besides the \textit{weight enumerator}, we are also interested in \textit{complete weight enumerator} of a linear code $\mathscr{C}$ over $E$ which is denoted by $cwe_{\mathscr{C}} ((X_i)_{i \in E})$ and is defined by, $cwe_{\mathscr{C}} ((X_i)_{i \in E})=\sum\limits_{\c \in \mathscr{C}} \prod\limits_{i \in E} X_i^{n_i(\c)}$ where $n_i(\c)$ registers the number of positions containing the value $i \in E$ in $\c$.    

\begin{note}
 The \textit{complete weight enumerator} of a $E$-linear code $\mathscr{C}$ of length $n$ is a homogeneous polynomial in $4$ indeterminate $X_0, X_a, X_b$ and $X_c$ and of degree $n$.
\end{note}

To provide a rigorous classification of the $E$-linear codes, we included the notion of \textit{Optimal code} as defined as the code $\mathscr{C}$ of length $n$ and of type $\{k_0,k_1\}$ to the fact that $\mathscr{C}$ possesses the largest minimum distance $d_{min}$ among all the codes that can constructed with the key parameters $n,k_0$ and $k_1$.  
We denote the largest minimum distance $d_{min}$ that can be obtained for given $n,k_0$ and $k_1$ by max$(d_{min})$.

Interestingly we can characterize linear codes $\mathscr{C}$ of length $n$ over $E$ by introducing two binary codes, the torsion and residue codes of length $n$.
The connected binary codes are defined as:

\begin{itemize}
    \item[$\bullet$] the residue code, $res(\mathscr{C})= \{\tau(\u) | \u \in \mathscr{C}\}$,
    \item[$\bullet$] the torsion code, $tor(\mathscr{C})= \{\v \in \mathbb{F}_2^n | c\v \in \mathscr{C}\}$.
\end{itemize}

As presented in \cite{ref3}, we denote dim$(res(\mathscr{C}))=k_0$ and dim$(tor(\mathscr{C}))-k_0=k_1$ which characterize the structure of the generating matrix for a linear code $\mathscr{C}$ of type $\{k_0,k_1\}$ from the fact that $|\mathscr{C}|=4^{k_0}.2^{k_1}$.  
Therefore, any linear code $\mathscr{C}$ of type $\{k_0,k_1\}$ admits a generating matrix $G$ over $E$ of the form 

\begin{equation}
 G=\left(
\begin{array}{ccc}
    aI_{k_0} & A & B  \\
    0 & cI_{k_1} & cC  \\
\end{array} \right)
\end{equation}

where $I_k$ implies the identity matrix of order $k$, $C$ is a binary matrix, whereas $A$ and $B$ are matrices over $E$.
The structure of $G$ shows that $k_1=0$ iff $\mathscr{C}$ is a free module. 

Unlike the binary case, any two $E$-codes $\mathscr{C}_1$ and $\mathscr{C}_2$ will be permutation equivalent iff one can find a coordinate permutation, which maps $\mathscr{C}_1$ to $\mathscr{C}_2$. 
To provide an alternate condition for the equivalency to take into account, Shi, Minjia, et al. \cite{ref9} stated that every permutation equivalent copy of a linear $E$-code $\mathscr{C}$ could be found from all possible coordinate relabeling in the $\mathbb{Z}_2$-span of its generator matrix of the form

\begin{equation}
 \left(
\begin{array}{ccc}
    aI_{k_0} & aT & aU  \\
    bI_{k_0} & bT & bU  \\
    0 & cI_{k_1} & cV  \\
\end{array} \right)
\label{generating matrix}
\end{equation}

Where $T, U$, and $V$ all are binary matrices.
For the rest of the paper, we use the above-said form as the generating matrix for an $E$-linear code to eliminate the permutation equivalent codes to appear in the classification, which we propose in the next section.
In some instances, the dimension of a code $\mathscr{C}$ over $E$ of type $\{k_0,k_1)\}$ can also be seen as $k=k_0+\frac{k_1}{2}$. 

The inner product over $E^n$ is defined by $\langle \u, \v \rangle =  \sum\limits_{i=1}^n u_i v_i$, for two codewords $\u=(u_1 \ u_2 \ \ldots \ u_n)$ and $\v=(v_1 \ v_2 \ \ldots \ v_n)$.
However, the non-commutativity of $E$ bifurcates the understanding of duality over $E^n$.   
The right dual code $\mathscr{C}^{\perp_{R}}$ of $\mathscr{C}$ is the right module $\mathscr{C}^{\perp_{R}} = \{\v \in E^n\ |\ \forall \u \in \mathscr{C}, \langle \u, \v \rangle = 0\}$, likewise the left dual code  is $\mathscr{C}^{\perp_{L}} = \{\v \in E^n\ |\ \forall \u \in \mathscr{C}, \langle \v, \u \rangle = 0\}$.

Consequently, the notion of self-orthogonal and self-dual code over $E$ has been introduced in \cite{ref3} as a code $\mathscr{C}$ over $E$ is defined to be left (resp. right) self-dual if it is equal to its left (resp. right) dual and hence $\mathscr{C}$ is self-dual if $\mathscr{C}= \mathscr{C}^{\perp_{L}} = \mathscr{C}^{\perp_{R}}$.
In similar way, a code $\mathscr{C}$ of length $n$ is said to be self-orthogonal if $\forall\ \u, \v \in \mathscr{C}, \langle \u, \v \rangle = 0$.
It is noteworthy that a code $\mathscr{C}$ over $E$ is self-orthogonal iff $\mathscr{C} \subseteq \mathscr{C}^{\perp_{L}}$ and  $\mathscr{C} \subseteq \mathscr{C}^{\perp_{R}}$, which simultaneously produces the necessary and sufficient condition for a code $\mathscr{C}$ over $E$ to be self-orthogonal as $\mathscr{C} \subseteq \mathscr{C}^{\perp_{L}} \cap \mathscr{C}^{\perp_{R}}$.

As an immediate consequence of the duality, one can make an observation rightfully pointed out in \cite{ref3}. 
We consider the following example for better understanding.

\begin{example}
 Consider the code $\mathscr{C}=\{0,a,b,c\}$ of length $1$ over $E$. 
 Now, $\mathscr{C}^{\perp_{R}}=\{0,c\}$ while $\mathscr{C}^{\perp_{L}}=\{0\}$.
This implies, $|\mathscr{C}|.|\mathscr{C}^{\perp_{R}}|=4.2=8 \neq 4$. 
\label{ex}
\end{example}

From Example \ref{ex}, it can be observed that the structure of codes constructed over $E$ differ from the conventional Coding Theory techniques as, in general, if $\mathscr{C}$ is a code of length $n$ with its dual $\mathscr{C}^{\perp}$ over a commutative ring $\mathcal{R}$ with unity then, $|\mathscr{C}|.|\mathscr{C}^{\perp}|= |\mathcal{R}|^n$. 

To overcome the obstacle, the concept of nice codes is introduced in \cite{ref3} that a code $\mathscr{C}$ of length $n$ over $E$ will be called left nice if $|\mathscr{C}|.|\mathscr{C}^{\perp_{L}}| = 4^n$.
The notion of right nice codes over $E$ can be obtained equivalently.
The code $\mathscr{C}$ is nice if it is both left and right nice.

In the following section, we classify the $E$-linear codes of length $\leq7$ that include the notion of optimality in terms of distance using a set of algorithms. 


\section{Classification of optimal $E$-linear codes of small length}
\label{classification}

So far, researchers' focus tends to be concentrated on the different construction of codes over $E$. 
Nonetheless, the crucial aspect of classifying the codes over $E$ is still lacking.
Since the classification undeniably impacted the case of codes over $\mathbb{Z}_4$ \cite{ref11, ref12}, we believe this work will facilitate the theory productively.

As we can ensure that all the $E$-linear codes of type $\{k_0,k_1\}$ that have the generator matrix of the form given in Expression \ref{generating matrix}, provide inequivalent codes, a simple calculation shows that there are precisely $2^{(k_0\times k_1)+(k_0\times (n-k_0-k_1))+(k_1\times (n-k_0-k_1)}$ codes are possible. 
Eventually, to reach the proposed outcome, we execute several algorithms to device different classification components mentioned above.

Algorithm \ref{alg:cap} is given below to construct the binary matrices $T, U$ and $V$ in the generating matrix $G$ given in Expression \ref{generating matrix} for linear codes $\mathscr{C}$ of type $\{k_0,k_1\}$. 

\begin{breakablealgorithm}
\caption{generate any Binary matrix $P$}\label{alg:cap}
\begin{algorithmic}[1]
\REQUIRE{$k$, value of an element $v$, $range$, dimension of the matrix $NP$}
\IF{$k=0$}
    \STATE return
\ENDIF
\\
\IF{$v=0$}
    \STATE $i=0$
    \WHILE{$i<2^{NP}$}
       \STATE $prev_i=i$
       \FOR{$i=i$ to $prev_i+range$}
            \STATE $P[i][NP-k]=v$
        \ENDFOR     
       \STATE $i=i+range$
    \ENDWHILE
\ENDIF
\\
\IF{$v=1$}
    \STATE $i=range$
    \WHILE{$i<2^{NP}$}
       \STATE $prev_i=i$
       \FOR{$i=i$ to $prev_i+range$}
            \STATE $P[i][NP-k]=v$
        \ENDFOR     
       \STATE $i=i+range$
    \ENDWHILE
\ENDIF
\\
\STATE $range=range/2$
\STATE $P(k-1,v,range,NP)$ 
\end{algorithmic}
\end{breakablealgorithm}

As the next step, we feed the constructed binary matrices $T, U$ and $V$ obtained from Algorithm \ref{alg:cap} in creating the generating matrices $G$ in Algorithm \ref{alg:cap1}.
One can note that the special treatment to identify the equivalent codes can be avoided by considering the form of the generating matrices as given in Expression \ref{generating matrix}. 

\begin{breakablealgorithm}
\caption{generator matrix}\label{alg:cap1}
\begin{algorithmic}[1]
\REQUIRE Number of rows $2k_0+k_1$, Number of columns $n$, dimension of the matrices $T, U, V$, respectively $NT, NU, NV$
\ENSURE $k_0+k_1< n$

\STATE generate $T(NT,0,2^{NT-1},NT)$ 
\STATE generate $T(NT,1,2^{NT-1},NT)$ 

\STATE generate $U(NU,0,2^{NU-1},NU)$ 
\STATE generate $U(NU,1,2^{NU-1},NU)$ 

\STATE generate $V(NV,0,2^{NV-1},NV)$ 
\STATE generate $V(NV,1,2^{NV-1},NV)$

\FOR{each possible matrix $T$}
  \STATE\FOR{each possible matrix $U$}
     \STATE\FOR{each possible matrix $V$}
            \STATE\FOR{$row=1$ to $k_0$}
               \STATE\FOR{$column=1$ to $k_1$}
                     \STATE Place an identity matrix multiplied by $a$
                      \ENDFOR
                \STATE\FOR{$column=k_0+1$ to $k_0+k_1$}    
                          \STATE Place generated binary matrix $T$ multiplied by $a$
                      \ENDFOR
                \STATE\FOR{$column=k_0+k_1+1$ to $n$} 
                          \STATE Place generated binary matrix $U$ multiplied by $a$
                      \ENDFOR
                 \ENDFOR
                 
            \STATE\FOR{$row=k_0+1$ to $k_0+k_1$}
               \STATE\FOR{$column=1$ to $k_0$}
                     \STATE Place an identity matrix multiplied by $b$
                      \ENDFOR
                \STATE\FOR{$column=k_0+1$ to $k_0+k_1$}    
                          \STATE Place generated binary matrix $T$ multiplied by $b$
                      \ENDFOR
                \STATE\FOR{$column=k_0+k_1+1$ to $n$} 
                          \STATE Place generated binary matrix $U$ multiplied by $b$
                      \ENDFOR
                 \ENDFOR     
          
            \STATE\FOR{$row=k_0+k_1+1$ to $2k_0+k_1$}
                    \STATE\FOR{$column=1$ to $k_0$}
                            \STATE Place a zero matrix
                         \ENDFOR
                   \STATE\FOR{$column=k_0+1$ to $k_0+k_1$}
                            \STATE Place an identity matrix multiplied by $c$
                         \ENDFOR
                    \STATE\FOR{$column=k_0+k_1+1$ to $n$}
                             \STATE Place generated binary matrix $V$ multiplied by $c$
                          \ENDFOR
                  \ENDFOR
                  \STATE Store Generator matrix $G$
        \ENDFOR
    \ENDFOR        
\ENDFOR
\end{algorithmic}
\end{breakablealgorithm}

Algorithm \ref{alg:cap2} deals with devising the codewords of code whose generating matrix can be obtained from Algorithm \ref{alg:cap1}.
Also, it is noteworthy that by deploying the form of the generating matrix of a code given in Expression \ref{generating matrix}, each codeword will lie in the $\mathbb{Z}_2$-sapn of $G$.   

\begin{breakablealgorithm}
\caption{Codewords generation}\label{alg:cap2}
\begin{algorithmic}[1]
\REQUIRE Generator matrix $G$ with number of rows and columns, $\alpha$ chosen from $\mathbb{Z}_2$
\STATE\FOR{each possible combination of $\alpha$ }
          \STATE\FOR{$k=1$ to $number\,of\,rows$}
                 \STATE multiply each $\alpha$ from the combination with $k^{th}$ row of Generator matrix $G$
                \ENDFOR
                \\
                \STATE add all the resultant rows of $G$
                \STATE return Codeword
                \\
      \ENDFOR
\end{algorithmic}
\end{breakablealgorithm}

The last crucial component in the schedule is depicted via Algorithm \ref{alg:cap3} to determine the minimum distance $d_{min}$ of a code $\mathscr{C}$ which can be extracted as the output of Algorithm \ref{alg:cap2}.    

\begin{algorithm}
\caption{Minimum Distance $d_{min}$}\label{alg:cap3}
\begin{algorithmic}[1]
\REQUIRE All codewords of a generator matrix $G$, Length of a codeword $n$
\\
\STATE $d_{min}=n$
\STATE\FOR{each codeword}
\STATE Compare codeword with every other codeword
\STATE count the number of positions that differ in both the compared codewords
\STATE \IF{$count<d_{min}$}
        \STATE $d_{min}=count$
        \ENDIF
      \ENDFOR
     \STATE return $d_{min}$
\end{algorithmic}
\end{algorithm}

To depict a near-to-complete or all-around classification, we listed the optimal $E$-linear codes of type $\{k_0,k_1\}$ in Table \ref{table 1} and in Table \ref{table 2}, which deal with codes of length $n\leq6$ and of length $n=7$ respectively.
With the given context, we denote $M(n,k_0,k_1)$ and $N(n,k_0,k_1)$ as the number of optimal codes and the number of codes that possesses the nice property respectively that can be found for given $n, k_0$ and $k_1$. 
Also, one can notice that the inclusion of codes of type $\{0,0\}$ for length $n\geq2$ is avoided in Table \ref{table 1} and Table \ref{table 2}. 

To exclude the possibility of the space criterion becoming an inevitable factor, we present the extensive amount of generated data for each of the codes for a specific set of parameters $n,k_0$ and $k_1$ in \url{https://guptalab.org/codesnur4/}, containing the generating matrix, codewords, codewords of the dual code, minimum distance, weight enumerator, complete weight enumerator and the criterion to state that whether the considered is a nice code.
On a particular note, the complexity of the algorithms and mechanism of the data visualization prevents a collective representation of the generated data for $n=7$. 
Hence we are forced to confine the information to the max$(d_{min})$ and the number of optimal codes for given $k_0$ and $k_1$, where $n=7$. 
In essence, we achieve from this mechanism, $M'(1)=0, M'(2)=4, M'(3)=24, M'(4)=160, M'(5)=1,472, M'(6)=20,096, M'(7)=420,096$ and $N'(1)=0, N'(2)=1, N'(3)=2, N'(4)=15, N'(5)=104, N'(6)=761$, where $M'(n)$ and $N'(n)$ stand for the number of inequivalent optimal codes and inequivalent nice codes for a given length $n$ over $E$ respectively by considering the Hamming distance.
It is worth noting that a higher number of optimal codes can be achieved in the view of the Hamming weight over $E$, compared to the same over the ring $\mathbb{Z}_4$ \cite{ref11,ref12}, in most of the instances for small lengths.

\begin{table}[h]
    \smaller
    \begin{tabular}{|c|c|c|c|c|c|}
    \hline
\multicolumn{1}{|c|}{$n$}          & \multicolumn{1}{c|}{$\{k_0, k_1\}$} & \multicolumn{1}{c|}{max$(d_{min})$} & \multicolumn{1}{c|}{$M(n,k_0,k_1)$} & Nice & $N(n,k_0,k_1)$  \\
\multicolumn{1}{|c|}{}          & \multicolumn{1}{c|}{} & \multicolumn{1}{c|}{} & \multicolumn{1}{c|}{} & Property &   \\
\hline
\hline
\multicolumn{1}{|c|}{$1$}                   & \multicolumn{1}{c|}{$\{0,0\}$}                 & \multicolumn{1}{c|}{-}          & \multicolumn{1}{c|}{-}          & \multicolumn{1}{c|}{-}          & \multicolumn{1}{c|}{-}                       \\ 
\hline
\hline

\multicolumn{1}{|c|}{\multirow{2}{*}{$2$}}  & \multicolumn{1}{c|}{$\{1,0\}$}                 & \multicolumn{1}{c|}{$2$}          & \multicolumn{1}{c|}{$1$}          & \multicolumn{1}{c|}{Yes}          & \multicolumn{1}{c|}{$1$}                       \\ \cline{2-6} 
\multicolumn{1}{|c|}{}                    & \multicolumn{1}{c|}{$\{0,1\}$}                 & \multicolumn{1}{c|}{$2$}          & \multicolumn{1}{c|}{$1$}          & \multicolumn{1}{c|}{No}          & \multicolumn{1}{c|}{-}                       \\ 
\hline
\hline

\multicolumn{1}{|c|}{\multirow{5}{*}{$3$}}  & \multicolumn{1}{c|}{$\{2,0\}$}                 & \multicolumn{1}{c|}{$2$}          & \multicolumn{1}{c|}{$1$}          & \multicolumn{1}{c|}{Yes}          & \multicolumn{1}{c|}{$1$}                       \\ \cline{2-6} 
\multicolumn{1}{|c|}{}                    & \multicolumn{1}{c|}{$\{1,0\}$}                 & \multicolumn{1}{c|}{$3$}          & \multicolumn{1}{c|}{$1$}          & \multicolumn{1}{c|}{Yes}          & \multicolumn{1}{c|}{$1$}                       \\ \cline{2-6} 
\multicolumn{1}{|c|}{}                    & \multicolumn{1}{c|}{$\{1,1\}$}                 & \multicolumn{1}{c|}{$2$}          & \multicolumn{1}{c|}{$2$}          & \multicolumn{1}{c|}{No}          & \multicolumn{1}{c|}{-}                       \\ \cline{2-6} 
\multicolumn{1}{|c|}{}                    & \multicolumn{1}{c|}{$\{0,1\}$}                 & \multicolumn{1}{c|}{$3$}          & \multicolumn{1}{c|}{$1$}          & \multicolumn{1}{c|}{No}          & \multicolumn{1}{c|}{-}                       \\ \cline{2-6} 
\multicolumn{1}{|c|}{}                    & \multicolumn{1}{c|}{$\{0,2\}$}                 & \multicolumn{1}{c|}{$2$}          & \multicolumn{1}{c|}{$1$}          & \multicolumn{1}{c|}{No}          & \multicolumn{1}{c|}{-}                       \\ 
\hline
\hline

\multicolumn{1}{|c|}{\multirow{9}{*}{$4$}}  & \multicolumn{1}{c|}{$\{3,0\}$}                 & \multicolumn{1}{c|}{$2$}          & \multicolumn{1}{c|}{$1$}          & \multicolumn{1}{c|}{Yes}          & \multicolumn{1}{c|}{$1$}                       \\ \cline{2-6} 
\multicolumn{1}{|c|}{}                    & \multicolumn{1}{c|}{$\{2,0\}$}                 & \multicolumn{1}{c|}{$2$}          & \multicolumn{1}{c|}{$9$}          & \multicolumn{1}{c|}{Yes}          & \multicolumn{1}{c|}{$1$}                       \\ \cline{2-6} 
\multicolumn{1}{|c|}{}                    & \multicolumn{1}{c|}{$\{2,1\}$}                 & \multicolumn{1}{c|}{$2$}          & \multicolumn{1}{c|}{$4$}          & \multicolumn{1}{c|}{No}          & \multicolumn{1}{c|}{-}                       \\ \cline{2-6} 
 \multicolumn{1}{|c|}{}                    & \multicolumn{1}{c|}{$\{1,0\}$}                 & \multicolumn{1}{c|}{$4$}          & \multicolumn{1}{c|}{$1$}          & \multicolumn{1}{c|}{Yes}          & \multicolumn{1}{c|}{$1$}                       \\ \cline{2-6}
\multicolumn{1}{|c|}{}                    & \multicolumn{1}{c|}{$\{1,1\}$}                 & \multicolumn{1}{c|}{$2$}          & \multicolumn{1}{c|}{$18$}          & \multicolumn{1}{c|}{No}          & \multicolumn{1}{c|}{-}                      \\ \cline{2-6}
\multicolumn{1}{|c|}{}                    & \multicolumn{1}{c|}{$\{1,2\}$}                 & \multicolumn{1}{c|}{$2$}          & \multicolumn{1}{c|}{$4$}          & \multicolumn{1}{c|}{Yes}          & \multicolumn{1}{c|}{$12$}                       \\ \cline{2-6}
\multicolumn{1}{|c|}{}                    & \multicolumn{1}{c|}{$\{0,1\}$}                 & \multicolumn{1}{c|}{$4$}          & \multicolumn{1}{c|}{$1$}          & \multicolumn{1}{c|}{No}          & \multicolumn{1}{c|}{-}                       \\ \cline{2-6} 
\multicolumn{1}{|c|}{}                    & \multicolumn{1}{c|}{$\{0,2\}$}                 & \multicolumn{1}{c|}{$2$}          & \multicolumn{1}{c|}{$9$}          & \multicolumn{1}{c|}{No}          & \multicolumn{1}{c|}{-}                       \\ \cline{2-6} 
\multicolumn{1}{|c|}{}                    & \multicolumn{1}{c|}{$\{0,3\}$}                 & \multicolumn{1}{c|}{$2$}          & \multicolumn{1}{c|}{$1$}          & \multicolumn{1}{c|}{No}          & \multicolumn{1}{c|}{-}                       \\ 
\hline
\hline

\multicolumn{1}{|c|}{\multirow{14}{*}{$5$}} & \multicolumn{1}{c|}{$\{4,0\}$}                 & \multicolumn{1}{c|}{$2$}          & \multicolumn{1}{c|}{$1$}          & \multicolumn{1}{c|}{Yes}          & \multicolumn{1}{c|}{$1$}                       \\ \cline{2-6} 
\multicolumn{1}{|c|}{}                    & \multicolumn{1}{c|}{$\{3,0\}$}                 & \multicolumn{1}{c|}{$2$}          & \multicolumn{1}{c|}{$27$}          & \multicolumn{1}{c|}{Yes}          & \multicolumn{1}{c|}{$1$}                      \\ \cline{2-6} 
\multicolumn{1}{|c|}{}                    & \multicolumn{1}{c|}{$\{3,1\}$}                 & \multicolumn{1}{c|}{$2$}          & \multicolumn{1}{c|}{$8$}          & \multicolumn{1}{c|}{No}          & \multicolumn{1}{c|}{-}                       \\ \cline{2-6} 
\multicolumn{1}{|c|}{}                    & \multicolumn{1}{c|}{$\{2,0\}$}                 & \multicolumn{1}{c|}{$3$}          & \multicolumn{1}{c|}{$12$}          & \multicolumn{1}{c|}{Yes}          & \multicolumn{1}{c|}{$1$}                      \\ \cline{2-6} 
\multicolumn{1}{|c|}{}                    & \multicolumn{1}{c|}{$\{2,1\}$}                 & \multicolumn{1}{c|}{$2$}          & \multicolumn{1}{c|}{$108$}          & \multicolumn{1}{c|}{No}          & \multicolumn{1}{c|}{-}                     \\ \cline{2-6} 
\multicolumn{1}{|c|}{}                    & \multicolumn{1}{c|}{$\{2,2\}$}                 & \multicolumn{1}{c|}{$2$}          & \multicolumn{1}{c|}{$15$}          & \multicolumn{1}{c|}{Yes}          & \multicolumn{1}{c|}{$36$}                      \\ \cline{2-6} 
\multicolumn{1}{|c|}{}                    & \multicolumn{1}{c|}{$\{1,0\}$}                 & \multicolumn{1}{c|}{$5$}          & \multicolumn{1}{c|}{$1$}          & \multicolumn{1}{c|}{Yes}          & \multicolumn{1}{c|}{$1$}                       \\ \cline{2-6} 
\multicolumn{1}{|c|}{}                    & \multicolumn{1}{c|}{$\{1,1\}$}                 & \multicolumn{1}{c|}{$3$}          & \multicolumn{1}{c|}{$24$}          & \multicolumn{1}{c|}{No}          & \multicolumn{1}{c|}{-}                      \\ \cline{2-6} 
\multicolumn{1}{|c|}{}                    & \multicolumn{1}{c|}{$\{1,2\}$}                 & \multicolumn{1}{c|}{$2$}          & \multicolumn{1}{c|}{$108$}          & \multicolumn{1}{c|}{Yes}          & \multicolumn{1}{c|}{$64$}                     \\ \cline{2-6} 
\multicolumn{1}{|c|}{}                    & \multicolumn{1}{c|}{$\{1,3\}$}                 & \multicolumn{1}{c|}{$2$}          & \multicolumn{1}{c|}{$8$}          & \multicolumn{1}{c|}{No}          & \multicolumn{1}{c|}{-}                       \\ \cline{2-6} 
\multicolumn{1}{|c|}{}                    & \multicolumn{1}{c|}{$\{0,1\}$}                 & \multicolumn{1}{c|}{$5$}          & \multicolumn{1}{c|}{$1$}          & \multicolumn{1}{c|}{No}          & \multicolumn{1}{c|}{-}                       \\ \cline{2-6} 
\multicolumn{1}{|c|}{}                    & \multicolumn{1}{c|}{$\{0,2\}$}                 & \multicolumn{1}{c|}{$3$}          & \multicolumn{1}{c|}{$12$}          & \multicolumn{1}{c|}{No}          & \multicolumn{1}{c|}{-}                      \\ \cline{2-6} 
\multicolumn{1}{|c|}{}                    & \multicolumn{1}{c|}{$\{0,3\}$}                 & \multicolumn{1}{c|}{$2$}          & \multicolumn{1}{c|}{$27$}          & \multicolumn{1}{c|}{No}          & \multicolumn{1}{c|}{-}                      \\ \cline{2-6} 
\multicolumn{1}{|c|}{}                    & \multicolumn{1}{c|}{$\{0,4\}$}                 & \multicolumn{1}{c|}{$2$}          & \multicolumn{1}{c|}{$1$}          & \multicolumn{1}{c|}{No}          & \multicolumn{1}{c|}{-}                       \\ 
\hline
\hline

\multicolumn{1}{|c|}{\multirow{20}{*}{$6$}} & \multicolumn{1}{c|}{$\{5,0\}$}                 & \multicolumn{1}{c|}{$2$}          & \multicolumn{1}{c|}{$1$}          & \multicolumn{1}{c|}{Yes}          & \multicolumn{1}{c|}{$1$}                       \\ \cline{2-6} 
\multicolumn{1}{|c|}{}                    & \multicolumn{1}{c|}{$\{4,0\}$}                 & \multicolumn{1}{c|}{$2$}          & \multicolumn{1}{c|}{$81$}          & \multicolumn{1}{c|}{Yes}          & \multicolumn{1}{c|}{$1$}                      \\ \cline{2-6} 
\multicolumn{1}{|c|}{}                    & \multicolumn{1}{c|}{$\{4,1\}$}                 & \multicolumn{1}{c|}{$2$}          & \multicolumn{1}{c|}{$16$}          & \multicolumn{1}{c|}{No}          & \multicolumn{1}{c|}{-}                       \\ \cline{2-6} 
\multicolumn{1}{|c|}{}                    & \multicolumn{1}{c|}{$\{3,0\}$}                 & \multicolumn{1}{c|}{$3$}          & \multicolumn{1}{c|}{$24$}          & \multicolumn{1}{c|}{Yes}          & \multicolumn{1}{c|}{$1$}                      \\ \cline{2-6} 
\multicolumn{1}{|c|}{}                    & \multicolumn{1}{c|}{$\{3,1\}$}                 & \multicolumn{1}{c|}{$2$}          & \multicolumn{1}{c|}{$648$}          & \multicolumn{1}{c|}{No}          & \multicolumn{1}{c|}{-}                     \\ \cline{2-6} 
\multicolumn{1}{|c|}{}                    & \multicolumn{1}{c|}{$\{3,2\}$}                 & \multicolumn{1}{c|}{$2$}          & \multicolumn{1}{c|}{$64$}          & \multicolumn{1}{c|}{Yes}          & \multicolumn{1}{c|}{$84$}                      \\ \cline{2-6} 
\multicolumn{1}{|c|}{}                    & \multicolumn{1}{c|}{$\{2,0\}$}                 & \multicolumn{1}{c|}{$4$}          & \multicolumn{1}{c|}{$12$}          & \multicolumn{1}{c|}{Yes}          & \multicolumn{1}{c|}{$1$}                       \\ \cline{2-6} 
\multicolumn{1}{|c|}{}                    & \multicolumn{1}{c|}{$\{2,1\}$}                 & \multicolumn{1}{c|}{$3$}          & \multicolumn{1}{c|}{$96$}          & \multicolumn{1}{c|}{No}          & \multicolumn{1}{c|}{-}                      \\ \cline{2-6} 
\multicolumn{1}{|c|}{}                    & \multicolumn{1}{c|}{$\{2,2\}$}                 & \multicolumn{1}{c|}{$2$}          & \multicolumn{1}{c|}{$1296$}          & \multicolumn{1}{c|}{Yes}          & \multicolumn{1}{c|}{$192$}                     \\ \cline{2-6} 
\multicolumn{1}{|c|}{}                    & \multicolumn{1}{c|}{$\{2,3\}$}                 & \multicolumn{1}{c|}{$2$}          & \multicolumn{1}{c|}{$64$}          & \multicolumn{1}{c|}{No}          & \multicolumn{1}{c|}{-}                       \\ \cline{2-6} 
\multicolumn{1}{|c|}{}                    & \multicolumn{1}{c|}{$\{1,0\}$}                 & \multicolumn{1}{c|}{$6$}          & \multicolumn{1}{c|}{$1$}          & \multicolumn{1}{c|}{Yes}          & \multicolumn{1}{c|}{$1$}                       \\ \cline{2-6} 
\multicolumn{1}{|c|}{}                    & \multicolumn{1}{c|}{$\{1,1\}$}                 & \multicolumn{1}{c|}{$4$}          & \multicolumn{1}{c|}{$24$}          & \multicolumn{1}{c|}{No}          & \multicolumn{1}{c|}{-}                      \\ \cline{2-6} 
\multicolumn{1}{|c|}{}                    & \multicolumn{1}{c|}{$\{1,2\}$}                 & \multicolumn{1}{c|}{$3$}          & \multicolumn{1}{c|}{$96$}          & \multicolumn{1}{c|}{Yes}          & \multicolumn{1}{c|}{$320$}                      \\ \cline{2-6} 
\multicolumn{1}{|c|}{}                    & \multicolumn{1}{c|}{$\{1,3\}$}                 & \multicolumn{1}{c|}{$2$}          & \multicolumn{1}{c|}{$648$}          & \multicolumn{1}{c|}{No}          & \multicolumn{1}{c|}{-}                       \\ \cline{2-6} 
\multicolumn{1}{|c|}{}                    & \multicolumn{1}{c|}{$\{1,4\}$}                 & \multicolumn{1}{c|}{$2$}          & \multicolumn{1}{c|}{$16$}          & \multicolumn{1}{c|}{Yes}          & \multicolumn{1}{c|}{$160$}                      \\ \cline{2-6}
\multicolumn{1}{|c|}{}                    & \multicolumn{1}{c|}{$\{0,1\}$}                 & \multicolumn{1}{c|}{$6$}          & \multicolumn{1}{c|}{$1$}          & \multicolumn{1}{c|}{No}          & \multicolumn{1}{c|}{-}                      \\ \cline{2-6}
\multicolumn{1}{|c|}{}                    & \multicolumn{1}{c|}{$\{0,2\}$}                 & \multicolumn{1}{c|}{$4$}          & \multicolumn{1}{c|}{$12$}          & \multicolumn{1}{c|}{No}          & \multicolumn{1}{c|}{-}                      \\ \cline{2-6}
\multicolumn{1}{|c|}{}                    & \multicolumn{1}{c|}{$\{0,3\}$}                 & \multicolumn{1}{c|}{$3$}          & \multicolumn{1}{c|}{$24$}          & \multicolumn{1}{c|}{No}          & \multicolumn{1}{c|}{-}                      \\ \cline{2-6}
\multicolumn{1}{|c|}{}                    & \multicolumn{1}{c|}{$\{0,4\}$}                 & \multicolumn{1}{c|}{$2$}          & \multicolumn{1}{c|}{$81$}          & \multicolumn{1}{c|}{No}          & \multicolumn{1}{c|}{-}                      \\ \cline{2-6}
\multicolumn{1}{|c|}{}                    & \multicolumn{1}{c|}{$\{0,5\}$}                 & \multicolumn{1}{c|}{$2$}          & \multicolumn{1}{c|}{$1$}          & \multicolumn{1}{c|}{No}          & \multicolumn{1}{c|}{-}                       \\
\hline
\end{tabular}
    \caption{Classification of inequivalent optimal codes and nice codes of length $n\leq6$ for the given parameters $k_0$ and $k_1$}
    \label{table 1}
\end{table}

\begin{table}[h]
    \centering
    \begin{tabular}{|c|c|c|c|}
    \hline
\multicolumn{1}{|c|}{$n$}          & \multicolumn{1}{c|}{$\{k_0, k_1\}$} & \multicolumn{1}{c|}{max$(d_{min})$} & \multicolumn{1}{c|}{$M(n,k_0,k_1)$} \\ 
\hline
\hline
\multicolumn{1}{|c|}{\multirow{27}{*}{$7$}} & \multicolumn{1}{c|}{$\{6,0\}$}                 & \multicolumn{1}{c|}{$2$}          & \multicolumn{1}{c|}{$1$}                       \\ \cline{2-4} 
\multicolumn{1}{|c|}{}                    & \multicolumn{1}{c|}{$\{5,0\}$}                 & \multicolumn{1}{c|}{$2$}          & \multicolumn{1}{c|}{$243$}                      \\ \cline{2-4} 
\multicolumn{1}{|c|}{}                    & \multicolumn{1}{c|}{$\{5,1\}$}                 & \multicolumn{1}{c|}{$2$}          & \multicolumn{1}{c|}{$32$}                       \\ \cline{2-4} 
\multicolumn{1}{|c|}{}                    & \multicolumn{1}{c|}{$\{4,0\}$}                 & \multicolumn{1}{c|}{$3$}          & \multicolumn{1}{c|}{$24$}                      \\ \cline{2-4} 
\multicolumn{1}{|c|}{}                    & \multicolumn{1}{c|}{$\{4,1\}$}                 & \multicolumn{1}{c|}{$2$}          & \multicolumn{1}{c|}{$3888$}                     \\ \cline{2-4} 
\multicolumn{1}{|c|}{}                    & \multicolumn{1}{c|}{$\{4,2\}$}                 & \multicolumn{1}{c|}{$2$}          & \multicolumn{1}{c|}{$256$}                      \\ \cline{2-4} 
\multicolumn{1}{|c|}{}                    & \multicolumn{1}{c|}{$\{3,0\}$}                 & \multicolumn{1}{c|}{$4$}          & \multicolumn{1}{c|}{$24$}                       \\ \cline{2-4} 
\multicolumn{1}{|c|}{}                    & \multicolumn{1}{c|}{$\{3,1\}$}                 & \multicolumn{1}{c|}{$3$}          & \multicolumn{1}{c|}{$192$}                      \\ \cline{2-4} 
\multicolumn{1}{|c|}{}                    & \multicolumn{1}{c|}{$\{3,2\}$}                 & \multicolumn{1}{c|}{$2$}          & \multicolumn{1}{c|}{$15552$}                     \\ \cline{2-4} 
\multicolumn{1}{|c|}{}                    & \multicolumn{1}{c|}{$\{3,3\}$}                 & \multicolumn{1}{c|}{$2$}          & \multicolumn{1}{c|}{$512$}                       \\ \cline{2-4} 
\multicolumn{1}{|c|}{}                    & \multicolumn{1}{c|}{$\{2,0\}$}                 & \multicolumn{1}{c|}{$4$}          & \multicolumn{1}{c|}{$190$}                       \\ \cline{2-4} 
\multicolumn{1}{|c|}{}                    & \multicolumn{1}{c|}{$\{2,1\}$}                 & \multicolumn{1}{c|}{$4$}          & \multicolumn{1}{c|}{$96$}                      \\ \cline{2-4} 
\multicolumn{1}{|c|}{}                    & \multicolumn{1}{c|}{$\{2,2\}$}                 & \multicolumn{1}{c|}{$3$}          & \multicolumn{1}{c|}{$384$}                      \\ \cline{2-4} 
\multicolumn{1}{|c|}{}                    & \multicolumn{1}{c|}{$\{2,3\}$}                 & \multicolumn{1}{c|}{$2$}          & \multicolumn{1}{c|}{$15552$}                       \\ \cline{2-4} 
\multicolumn{1}{|c|}{}                    & \multicolumn{1}{c|}{$\{2,4\}$}                 & \multicolumn{1}{c|}{$2$}          & \multicolumn{1}{c|}{$256$}                      \\ \cline{2-4}
\multicolumn{1}{|c|}{}                    & \multicolumn{1}{c|}{$\{1,0\}$}                 & \multicolumn{1}{c|}{$7$}          & \multicolumn{1}{c|}{$1$}                      \\ \cline{2-4}
\multicolumn{1}{|c|}{}                    & \multicolumn{1}{c|}{$\{1,1\}$}                 & \multicolumn{1}{c|}{$4$}          & \multicolumn{1}{c|}{$380$}                      \\ \cline{2-4}
\multicolumn{1}{|c|}{}                    & \multicolumn{1}{c|}{$\{1,2\}$}                 & \multicolumn{1}{c|}{$4$}          & \multicolumn{1}{c|}{$96$}                      \\ \cline{2-4}
\multicolumn{1}{|c|}{}                    & \multicolumn{1}{c|}{$\{1,3\}$}                 & \multicolumn{1}{c|}{$3$}          & \multicolumn{1}{c|}{$192$}                      \\ \cline{2-4}
\multicolumn{1}{|c|}{}                    & \multicolumn{1}{c|}{$\{1,4\}$}                 & \multicolumn{1}{c|}{$2$}          & \multicolumn{1}{c|}{$3888$}                      \\ \cline{2-4}
\multicolumn{1}{|c|}{}                    & \multicolumn{1}{c|}{$\{1,5\}$}                 & \multicolumn{1}{c|}{$2$}          & \multicolumn{1}{c|}{$32$}                      \\ \cline{2-4}
\multicolumn{1}{|c|}{}                    & \multicolumn{1}{c|}{$\{0,1\}$}                 & \multicolumn{1}{c|}{$7$}          & \multicolumn{1}{c|}{$1$}                      \\ \cline{2-4}
\multicolumn{1}{|c|}{}                    & \multicolumn{1}{c|}{$\{0,2\}$}                 & \multicolumn{1}{c|}{$4$}          & \multicolumn{1}{c|}{$190$}                      \\ \cline{2-4}
\multicolumn{1}{|c|}{}                    & \multicolumn{1}{c|}{$\{0,3\}$}                 & \multicolumn{1}{c|}{$4$}          & \multicolumn{1}{c|}{$24$}                      \\ \cline{2-4}
\multicolumn{1}{|c|}{}                    & \multicolumn{1}{c|}{$\{0,4\}$}                 & \multicolumn{1}{c|}{$3$}          & \multicolumn{1}{c|}{$24$}                      \\ \cline{2-4}
\multicolumn{1}{|c|}{}                    & \multicolumn{1}{c|}{$\{0,5\}$}                 & \multicolumn{1}{c|}{$2$}          & \multicolumn{1}{c|}{$243$}                      \\ \cline{2-4}
\multicolumn{1}{|c|}{}                    & \multicolumn{1}{c|}{$\{0,6\}$}                 & \multicolumn{1}{c|}{$2$}          & \multicolumn{1}{c|}{$1$}                      \\
\hline
    \end{tabular}
    \caption{Classification of inequivalent optimal codes of length $n=7$ for the given parameters $k_0$ and $k_1$}
    \label{table 2}
\end{table}


\section{Conclusion and Future work}
\label{conclusion}
 This work centered around the classification of linear \textit{Optimal codes} of length $n\leq7$ over the non-commutative and non-unital ring $E$ of order $4$. 
 Most of the critical factors that play essential roles in the individual characterization and further analysis of $E$-linear codes of length $n$ and of type $\{k_0,k_1\}$, such as generating matrices, minimum distances, dual codewords of each of the codes, weight enumerators, and complete weight enumerators, are considered in this article.
 Moreover, for length $n\leq6$, the decisive argument of whether a linear code $\mathscr{C}$ is a nice code is conferred.
 The computational results can be visualized more effectively in \cite{ref13} for a given set of parameters $n,k_0$, and $k_1$ by deploying a set of algorithms.
 For future work, we would like to extend this work by optimizing the algorithms in terms of the time complexity and introducing a more robust classification of codes of length more than $7$.
 Furthermore, the possibilities of building cryptographic protocols over the ring $E$ using the \textit{Hull dimension} or by proposing lattice construction are still unexplored, which one should pursue in particular.
 
%


 





\end{document}